\documentclass{emulateapj}

\newcommand{\kms}{km s$^{-1}$}

\slugcomment{Last updated: \today}
\shorttitle{Orbital Dependence of Galaxy Properties}
\shortauthors{Hwang \& Park}

\begin{document}

\title{Orbital Dependence of Galaxy Properties in Satellite Systems of Galaxies}
\author{Ho Seong Hwang\altaffilmark{1,2} and Changbom Park\altaffilmark{1}}
\altaffiltext{1}{School of Physics, Korea Institute for Advanced Study, Seoul 130-722, Korea}
\altaffiltext{2}{CEA Saclay/Service d'Astrophysique, F-91191 Gif-sur-Yvette, France}
\email{hoseong.hwang@cea.fr, cbp@kias.re.kr}

\begin{abstract}

We study the dependence of satellite galaxy properties 
  on the distance to the host galaxy and the orbital motion (prograde and retrograde orbits)
  using the Sloan Digital Sky Survey (SDSS) data.
From SDSS Data Release 7 we find 3515 isolated satellite systems of galaxies at $z<0.03$ 
  that contain 8904 satellite galaxies.
Using this sample we construct a catalog of 635 satellites associated with 215 host galaxies
  whose spin directions are determined by our inspection of the SDSS color images 
  and/or by spectroscopic observations in the literature.
We divide satellite galaxies into prograde and retrograde orbit subsamples depending 
  on their orbital motion respect to the spin direction of the host.
We find that
  the number of galaxies in prograde orbit is nearly equal to that of retrograde orbit 
  galaxies: the fraction of satellites in prograde orbit is $50\pm2$\%.
The velocity distribution of satellites with respect to their hosts
  is found almost symmetric: the median bulk rotation of satellites is $-1\pm 8$ km s$^{-1}$.
It is found that the radial distribution of early-type satellites in prograde orbit is strongly 
  concentrated toward the host while that of retrograde ones shows much less concentration.
We also find the orbital speed of late-type satellites in prograde orbit increases 
 as the projected distance to the host ($R$) decreases
 while the speed decreases for those in retrograde orbit. 
At $R$ less than 0.1 times the host virial radius 
  $(R<0.1 r_{\rm vir,host})$ the orbital speed decreases in both prograde and 
  retrograde orbit cases.
Prograde satellites are on average fainter than retrograde satellites for both
  early and late morphological types.
The $u-r$ color becomes redder as $R$ decreases for both prograde and retrograde 
  orbit late-type satellites.
The differences between prograde and retrograde orbit satellite galaxies 
  may be attributed to their different origin or the different strength 
  of physical processes that they have experienced 
  through hydrodynamic interactions with their host galaxies.
\end{abstract}


\section{Introduction}

In the hierarchical picture of galaxy formation,
  massive galaxies are formed through merger and accretion.
Satellite galaxies associated with massive host galaxies,
 memorize the history of their accretion and interaction with host galaxies.
In addition, they are distributed out to a few virial radii of the host galaxy.
Therefore, they are excellent probes of the dark matter halos
  surrounding the host galaxies.

A great deal of effort has been made to understand the spatial distribution
  of satellite galaxies around their host galaxies.
The radial distribution of the observed satellites 
  was found to be more concentrated than that of subhalos 
  around galaxy-sized halos in N-body simulations \citep{chen06},
  but consistent with dark matter distribution \citep{van05,chen06}.
The studies of color dependence on the radial distribution of satellites
  revealed that the radial distribution of red satellites is 
  more centrally concentrated than that of blue satellites \citep{sal07,chen08}.
Similarly, the radial distribution of early morphological type satellites
  is more centrally concentrated than that of late-type satellites \citep{ann08,wang10}.
The angular distribution of satellite galaxies is also an issue,
which has been extensively addressed
  (e.g., \citealt{pra03,sl04,sl05,ann08,bai08,wang09}).

Another interesting issue is the kinematic properties of satellite galaxies.
Satellite galaxies can be divided into two classes based on their orbital motion:
  those in prograde orbit that 
  rotate around the host in the same direction as the rotation of the host,
  and those in retrograde orbit that rotate in the opposite direction.
Several important questions have been raised for these two classes: 
  (1) what the number ratio of galaxies in prograde orbit to those in retrograde orbit is;
  (2) what the amplitude of mean bulk rotation of satellites is; and
  (3) whether there is a difference in physical properties between them.

\citet{zar93} searched for the satellites around late-type spiral galaxies 
  with a luminosity similar to the Milky Way, and 
  found 69 satellites around 45 host galaxies ($\sim$1.5 satellites per host).
For 35 satellites whose orbital velocity directions were determined,
  they detected a small amount of rotation for the satellite systems, $\sim29\pm21$ km s$^{-1}$,
  in the same sense as the rotation of hosts' disk.
They also found that the number of satellites in prograde orbit
  is nearly equal to that in retrograde orbit.
However, in NGC 5084, \citet{car97} detected nine satellite galaxies,
  and found a clear excess of satellites in retrograde orbit
  (eight in retrograde orbit and one in prograde orbit).
Using the expanded catalog of 115 satellites around 69 host galaxies,
  \citet{zar97} found that the number of galaxies in prograde orbit 
  is almost the same as that of galaxies in retrograde orbit.

Recently, \citet{azz06} measured the direction of rotation 
  for some host galaxies in the list of satellite systems of \citet{pra03},
  and secured 76 satellites associated with 43 host galaxies
  of which orbital directions were determined.
With this sample,
  they found that the fraction of galaxies in prograde orbit ($f_{\rm pro}$) is $\sim0.6$, 
  which is slightly larger than the previous results ($f_{\rm pro}\sim0.52$ in \citealt{zar97}),
  but is broadly consistent with those from their $N$-body simulation ($f_{\rm pro}\sim0.55-0.60$).
In the catalog of satellite systems extracted from the Sloan Digital Sky Survey (SDSS; \citealt{york00}) 
  Data Release 6 \citep{bai08},
  \citet{hf08} determined the direction of rotation of some host galaxies
  by spectroscopic observations.
They obtained the data for 78 satellites associated with 63 hosts.
Combining these data with those for the satellites in \citet{zar97},
  they found that the fraction of galaxies in prograde orbit is still $0.53\pm0.04$, and
  mean bulk rotation of satellites is $37\pm3$ km s$^{-1}$,
  in the same sense as the rotation of hosts' disk.
Interestingly, they found that the peculiar velocity distribution of satellites
  is not a single Gaussian, but is described as the sum of two Gaussians.

As summarized above, most studies focused on
  the first two issues, the fraction of galaxies in prograde orbit and 
  the amplitude of mean bulk rotation of satellites.
The difference in physical properties between them,
  which is very important to understand the formation of satellite systems,
  has not been explored much.
Moreover, the direction of rotation of host galaxy that is 
  essential to know the orbit of satellites,
  has been determined by the expensive spectroscopic observations.

In this paper,
  we determine the direction of rotation of host galaxies
  using the color images provided by the SDSS
  without conducting spectroscopic observations (see \S \ref{spin} for details),
  and present the results of a study on the dependence of various physical properties
  of satellite galaxies on their orbits (prograde and retrograde orbits).
In this paper we mean by `satellites' as the galaxies that are much fainter than
  their host galaxies that are isolated from other bright galaxies.
Therefore, they are like the conventional satellites associated with the Milky Way or Andromeda,
  and are rather different from the `satellites' 
  used in the N-body simulations or group/cluster studies
  where all galaxies (or dark halos) other than 
  the brightest one (or the most massive one) are called satellites.
Section \ref{data} describes the observational data used in this study.
Orbital dependence of the physical parameters of satellites are given in \S \ref{results}.
Discussion and summary are given in \S \ref{discuss} and \S \ref{sum}, respectively.

\section{Observational Data Set}\label{data}
\subsection{Physical Parameters of Galaxies}

We use a spectroscopic sample of galaxies in the SDSS DR7 \citep{aba09}.
The physical parameters of galaxies that we consider in this study are
  $r$-band absolute Petrosian magnitude ($M_r$), morphology, ($u-r$) color,
  equivalent width of $H\alpha$ emission line [W(H$\alpha$)], color gradient in ($g-i$) color,
  concentration index ($c_{\rm in}$), internal velocity dispersion ($\sigma$), and
  Petrosian radius in $i$-band ($R_{\rm Pet}$).
Here we give a brief description of these parameters.

The $r$-band absolute magnitude $M_r$ was computed using the formula,
\begin{equation}\label{eq-mag}
M_r=m_r-5{\rm log}_{10}[r(z)(1+z)]-K(z)+E(z),
\end{equation}
where $r(z)$ is a comoving distance at redshift $z$,
  $K(z)$ is $K$-correction, and $E(z)$ is the luminosity evolution correction.
We adopt a flat $\Lambda$CDM cosmology with density parameters 
  $\Omega_\Lambda=0.73$ and $\Omega_m=0.27$.
The rest-frame absolute magnitudes of
  individual galaxies are computed in fixed bandpasses, shifted to $z=0.1$,
  using Galactic reddening correction \citep{sch98} and $K$-corrections
  as described by \citet{bla03}.
The evolution correction given by \citet{teg04}, $E(z) = 1.6(z-0.1)$, is also applied.
The distance $r(z)$ has a unit of $h^{-1}$Mpc and the corresponding $5{\rm log}h$ in $M_r$
  will be omitted in this paper.

We adopt the galaxy morphology information from 
  the SDSS DR7 release of the Korea Institute for Advanced Study value-added galaxy catalog 
  (KIAS-VAGC; Choi 2010, in prep.). 
In this catalog galaxies are divided into early (ellipticals and lenticulars) and
  late (spirals and irregulars) morphological types based on their locations
  in the ($u-r$) color versus ($g-i$) color gradient space and also in the
  $i$-band concentration index space \citep{pc05}.
The resulting morphological classification has completeness and reliability reaching 90\%.
We perform an additional visual check of the color images of galaxies 
  to correct misclassifications by the automated scheme.
In this procedure we changed the types of the
  blended or merging galaxies, blue but elliptical-shaped galaxies,
  and dusty edge-on spirals.
  
The $^{0.1}(u-r)$ color was computed using the extinction and $K$-corrected model magnitude.
The superscript 0.1 means the rest-frame magnitude $K$-corrected to the redshift of 0.1,
  and will subsequently be dropped.
We adopt the values of $(g-i)$ color, concentration index ($c_{\rm in}$),
  and Petrosian radius $R_{\rm Pet}$
  computed for the galaxies in KIAS-VAGC (Choi 2010, in prep.).
The $(g-i)$ color gradient was defined by the color difference
  between the region with $R_{\rm gal}<0.5R_{\rm Pet}$ and the annulus with
  $0.5R_{\rm Pet}<R_{\rm gal}<R_{\rm Pet}$,
  where $R_{\rm gal}$ is the galactocentric radius and
  $R_{\rm Pet}$ is the Petrosian radius estimated in the $i$-band image.
To account for the effects of flattening or inclination of galaxies,
  elliptical annuli were used to calculate the parameters.
The (inverse) concentration index is defined by $R_{50}/R_{90}$,
  where $R_{50}$ and $R_{90}$ are semimajor axis lengths of ellipses
  containing $50\%$ and $90\%$ of the Petrosian flux in the $i$-band image, respectively.

The velocity dispersion values of galaxies are adopted 
  from the DR7 release of New York University (NYU) VAGC \citep{bla05}
  and values of $H\alpha$ equivalent width are taken 
  from the DR7 release of 
  Max-Planck-Institute for Astrophysics (MPA)/John Hopkins University (JHU) VAGC \citep{tre04}.

\subsection{Satellite Systems of Isolated Galaxies}\label{isolate}

Finding isolated satellite systems consists of two steps:
  identifying isolated galaxies and finding satellites associated with the isolated galaxies.
We consider all SDSS galaxies with redshifts less than 0.03 as the host candidates.
Only the relatively nearby galaxies are considered because the visual determination
  of the sense of rotation is feasible only when the galaxy image shows
  detailed internal features.
For each galaxy its neighbors are searched.
The neighbors of a target galaxy with a $r$-band absolute magnitude $M_r$ are the galaxies
  with an absolute magnitude brighter than $M_r$+1 and a velocity difference less than $\Delta v$.
We adopt $\Delta v=1000$ km s$^{-1}$ for early-type target galaxies, 
  and $750$ km s$^{-1}$ for late-type galaxies based on the distributions of 
  pairwise velocity difference of the SDSS galaxies (see Fig. 2 of \citealt{pc09} and
  Fig. 1 of \citealt{wang09}).
The neighbors are searched in the redshift interval from $-0.005$ to $0.034$.
When a target galaxy has a projected distance to its neighbor galaxies ($R_{\rm nei}$) larger
  than max[$r_{\rm vir,target}$,$r_{\rm vir,nei}$] for all neighbors,
  it is selected as an isolated galaxy.
Here $r_{\rm vir}$ is the virial radius of a galaxy (see eq. (5) of \citealt{pc09}).

Once isolated host candidates are determined, the satellites associated with them
  are searched for.
A galaxy qualifies as a satellite if it has an absolute magnitude fainter than $M_r$(host)+1, 
  velocity difference less than $600$ km s$^{-1}$,
  and projected separation of the galaxy with the host
  $R<$min[$R_{\rm nei}$ $r_{\rm vir,host}$/($r_{\rm vir,host}+r_{\rm vir,nei}$),
  $R_{\rm nei}-r_{\rm vir,nei}$]
  where $R_{\rm nei}$ is a projected separation between the host and host's neighbor.
The first limit divides the distance between the host and neighbor in proportion
  to their virial radii. The second limit prohibits the galaxies within the virial
  radii of the neighbor galaxies from being selected as satellites.
In our SDSS galaxy sample we find 8904 satellites associated with 3515 isolated host galaxies.

\subsection{Spin Direction of Host Galaxies}\label{spin}

To determine the sense of rotation of host galaxies from their images
  we first assume that all spiral galaxies have trailing arms \citep{bt87}
  and determine the receding side of the major axis from their color images.
The part of minor axis behind the bulge often appears smooth due to the foreground bulge stars
  while the side of the minor axis closer than the bulge often shows
  interstellar dust patches of relatively higher sharpness \citep{mb81}.
When it is not possible to determine the sense of rotation due to various reasons,
  we simply drop the host galaxy.
Very early-type galaxies with smooth disks, very late-type galaxies with vanishing bulge,
  and nearly face-on or edge-on disk galaxies are difficult to
  assign the spin direction.
After we remove all these cumbersome cases, 
  we can determine the spin vector directions for 165 host galaxies
  that have 535 satellite galaxies.

We supplemented our sample by adding host galaxies of which rotation sense 
  was determined from the spectroscopic observations in the literature
  \citep{rub82,rub85,rub99,bro94,han95,tay95,rhee96,zar97,sof97p,sof97a,sof98,sof03,
  her98,her99,vb01,gar02,gar03,gar04,gar05,vogt04,coc04,fri05,her05,noo05,
  che06,fb06,dai06,gan06,azz06,hf08,epi08a,epi08b,swa09}.
We compiled 390 satellites associated with 116 hosts.
There are 51 hosts in common between the hosts of which rotation sense
  was determined in this study and those in the literature.
We have compared our determination of the rotation sense with that in the literature
  for these common objects, and found that there is only one galaxy (NGC 5894)
  that shows a disagreement between the two.
We kept the rotation sense for this object determined from the spectroscopic observation
  in the literature.
In total, we have 727 satellites assigned to 230 hosts whose spin vector directions are known.

Among 727 satellites,
  we eliminated 45 spurious sources (e.g., faint fragments of bright galaxies, 
  diffraction spikes of bright stars)
  that were included in the list of satellites
  (originally included in the spectroscopic sample of galaxies in SDSS)
  by inspecting color images of the satellites.
In the result, 682 satellites with 221 host galaxies remain.

It is noted that our list contains some satellite systems whose
  host galaxies are early types because their spin directions
  were determined from the spectroscopic observations.
Since the morphology of satellite galaxies is strongly affected by the
  morphology of host galaxies as found in \citet{ann08},
  we further divide the satellites into two categories (the cases of early- and late-type hosts)
  and obtain 197 late-type host galaxies having 579 satellite galaxies
  and 24 early-type hosts with 103 satellites.
Our analysis will be restricted to the satellite systems with late-type hosts except \S \ref{samp}.

\begin{figure}
\center
\includegraphics[width=85mm]{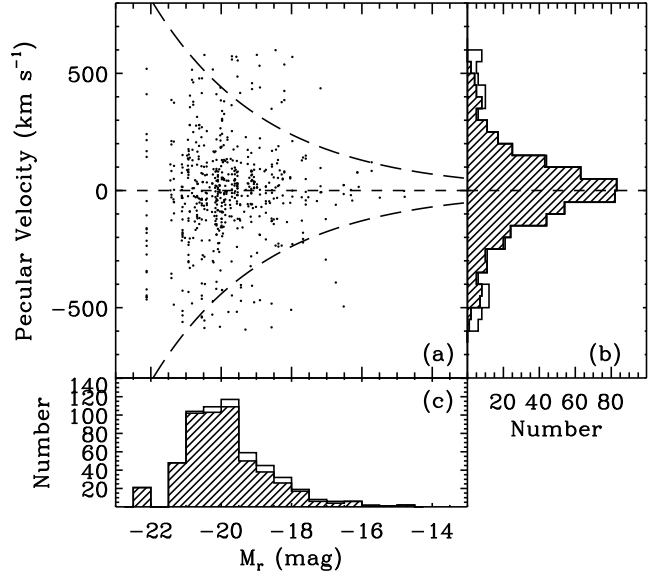}
\caption{({\it a}) Peculiar velocities of satellites 
  as a function of absolute magnitudes of late-type host galaxies.
Dashed-line envelops indicate the velocity criteria for rejecting interlopers. 
({\it b}) Distribution of absolute magnitudes of all hosts (open histogram)
  and of those having satellites within the velocity criteria (shaded histogram).
({\it c}) Distribution of peculiar velocities of all satellites (open histogram)
  and of those within the velocity criteria (shaded histogram). 
 }\label{fig-ncut}
\end{figure}

In Figure \ref{fig-ncut}, we plot peculiar velocities of satellite galaxies
  as a function of absolute magnitudes of host galaxies.
The peculiar velocity of a satellite is defined by, 
  $v_{\rm pec}=S(v_{\rm sat}-v_{\rm host})$, where $v_{\rm sat}$ and $v_{\rm host}$ 
  are observed velocities of the satellite and the host, respectively.
$S$ is defined by
\begin{eqnarray}
\label{eq-S}
S = \left\{ \begin{array}{l}
  +1,~\theta < 90^{\circ}, \\
  -1,~\mathrm{otherwise.} \end{array} \right. \nonumber
\end{eqnarray}

$\theta=\mid \mathrm{PAred} - \mathrm{PAS} \mid$
  where PAred is a position angle of the receding side of the host's major axis, 
  and PAS is the position angle of the satellite with respect to the host.
All position angles are measured from north to east.
Therefore, the satellites with $v_{\rm pec}>0$ 
  rotate around the host in the same direction as the rotation of the host (prograde orbit)
  and those with $v_{\rm pec}<0$ rotate in the opposite direction (retrograde orbit).

Since the probability for a galaxy to become a satellite of a host
  decreases as the mass (luminosity) of the host decreases,
  we exclude some probable interlopers 
  located outside the region bounded by dashed lines in Figure \ref{fig-ncut}.
The dashed lines are determined as follows.
The escape velocity of a satellite can be given by 
  $v_{\rm esc}\propto(M_{\rm host}/R)^{0.5}$
  where $M_{\rm host}$ is a mass of the host and 
  $R$ is a projected distance to the host.
The mass of the host galaxy is computed by $M_{\rm host}=\gamma L$ 
  where $\gamma$ is the mass-to-light ratio and $L$ is the $r$-band luminosity of the host.
We assume that $\gamma$(early)=$2\gamma$(late) at the same $r$-band luminosity,
  and that $\gamma$ is constant with galaxy luminosity 
  for a given morphological type \citep{park08}.
Then we obtain $v_{\rm esc}(M_r)=v_0 10^{(-0.4/3) (M_r-M_0)}$ 
  in a given distance to the host galaxy (e.g., $R=r_{\rm vir,host}$).
We adopt $v_0=600$ km s$^{-1}$ and $M_0=-21$ mag, and plot $v_{\rm esc}$ 
  versus $M_r$ as dashed lines in Figure \ref{fig-ncut}.

\begin{figure}
\center
\includegraphics[width=85mm]{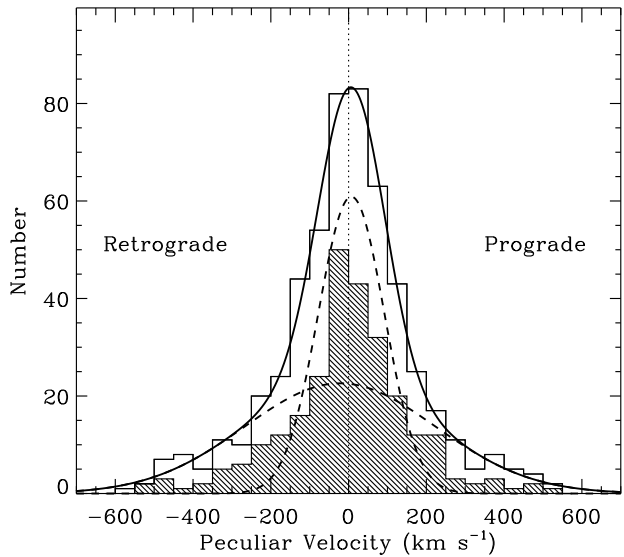}
\caption{Peculiar velocity distribution for all satellite galaxies
  associated with late-type host galaxies.
The vertical dotted line indicates zero velocity.
Dashed lines represent the double Gaussian fits, and the solid line indicates
  the sum of two Gaussian functions.
The hatched histogram indicates the peculiar velocity distribution 
  of satellite galaxies located near the major axis of the hosts 
  ($\theta < 45^{\circ}$).
 }\label{fig-hist}
\end{figure}

\begin{figure}
\center
\includegraphics[width=85mm]{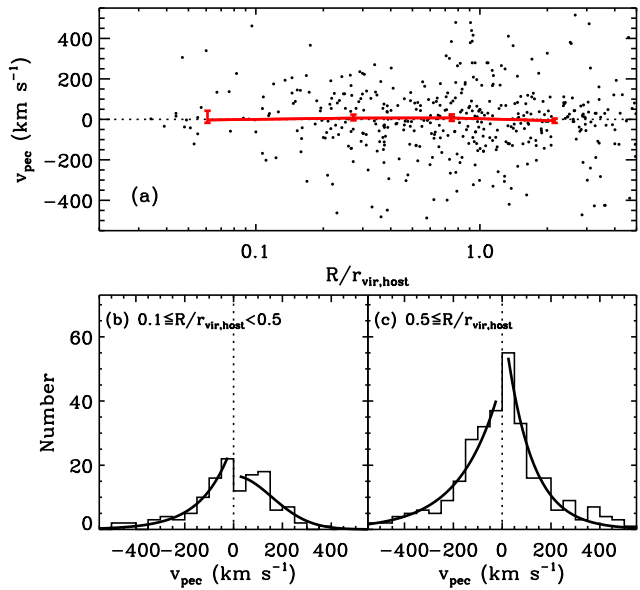}
\caption{(a) Peculiar velocity $v_{\rm pec}$ vs. projected distance to the host galaxy.
Median values in each radial bin and their errors are connected by solid lines.
Peculiar velocity distribution for the satellites 
  (b) at $0.1 \le R/r_{\rm vir,host} < 0.5$ and (c) at $0.5\le R/r_{\rm vir,host}$.
The vertical dotted lines indicate zero velocity.
Solid lines indicate the best fit exponential functions of the distributions of satellites 
  in prograde or retrograde orbits separately.
Solid line for the satellites in prograde orbit 
  at $0.1 \le R/r_{\rm vir,host} < 0.5$ is the best fit Gaussian function. 
}\label{fig-histrad}
\end{figure}

The virial radius of a host galaxy ($r_{\rm vir,host}$)
  is defined by the projected radius where the mean mass density $\rho$
  within the sphere of radius $r$ is 200 times the critical density
  or 740 times the mean density of the universe, namely,
\begin{equation}
r_{\rm vir} = (3 \gamma L /4\pi / 200{\rho_c})^{1/3}.
\end{equation}
Since we adopt $\Omega_m = 0.27$, $200\rho_c = 200 {\bar\rho}/\Omega_m = 740{\bar\rho}$
  where $\bar\rho$ is the mean density of the universe.
The value of mean density of the universe, 
  $\bar\rho=(0.0223\pm0.0005)(\gamma L)_{-20} (h^{-1}{\rm Mpc})^{-3}$,
  was adopted, where $(\gamma L)_{-20}$ is the mass of a late-type galaxy 
  with $M_r=-20$ \citep{park08}.
According to our formula the virial radii of galaxies with
  $M_r=-19.5,-20.0,$ and $-20.5$ are 260, 300, and 350 $h^{-1}$ kpc for early types,
  and 210, 240, and 280 $h^{-1}$ kpc for late types, respectively.
By rejecting probable interlopers as shown in Figure \ref{fig-ncut},
  we finally secured 534 satellite galaxies associated with 191 late-type hosts
  and 101 satellites associated with 24 early-type hosts.

We present the peculiar velocity distribution of 
  all 534 satellite galaxies in Figure \ref{fig-hist}.
The fraction of galaxies in prograde orbit ($f_{\rm pro}$) is found to be $50\pm2\%$, 
  and the median value of the peculiar velocity distribution is $-1\pm8$ km s$^{-1}$.
When we use only 262 satellites located near the major axis of the hosts 
  ($\theta < 45^{\circ}$),
  $f_{\rm pro}$ does not change 
  and the median value of the distribution is $0\pm10$ km s$^{-1}$.
We fit the peculiar velocity distribution using the sum of two Gaussian functions,
  and find the best-fit function with mean values of 
  $\left< v_{\rm pec} \right> =7$ \kms ($\sigma=85$ \kms) and
  $\left< v_{\rm pec} \right> =-22$ \kms($\sigma=246$ \kms).
When we fit the peculiar velocity distribution of satellites 
  in prograde or retrograde orbits separately using a single Gaussian or exponential function,
  we find that the exponential function fits the distribution better than the Gaussian function.
The reduced $\chi^2$ for satellites in prograde orbit, are 2.1 and 0.5
  for Gaussian and exponential functions, respectively.
The reduced $\chi^2$ for satellites in retrograde orbit are 2.5 and 0.7
  for Gaussian and exponential functions, respectively.

To study whether or not the peculiar velocity distribution changes 
  depending on the distance to the host
  we show the distribution for satellites at $0.1 \le R/r_{\rm vir,host} < 0.5$ or
  at $0.5\le R/r_{\rm vir,host}$ separately in Figure \ref{fig-histrad}.
We use Gaussian and exponential functions to fit the distributions 
  of satellites in prograde and retrograde orbits, and
  show best-fit functions in Figure \ref{fig-histrad}(b-c).
It is seen that the best-fit function for satellites in retrograde orbit 
  is exponential both in the inner and outer regions,
  while that for satellites in prograde orbit changes from Gaussian (inner region)
  to exponential (outer region).

\begin{figure}
\center
\includegraphics[width=85mm]{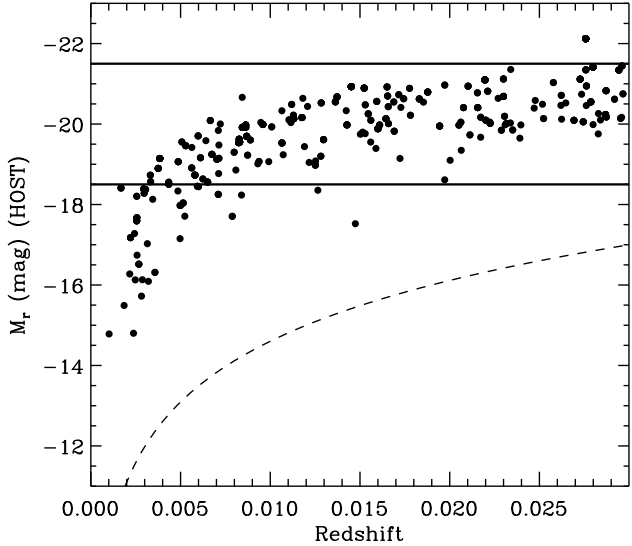}
\caption{Absolute magnitude vs. redshift for late-type host galaxies.
The dashed line is the apparent magnitude limit ($m_r=17.77$)
  computed from eq. (\ref{eq-mag}) in \citet{choi07}.
The satellites associated with the host galaxies within the solid box
  are selected as a luminosity subsample of satellites. 
}\label{fig-host}
\end{figure}

\begin{figure}
\center
\includegraphics[width=85mm]{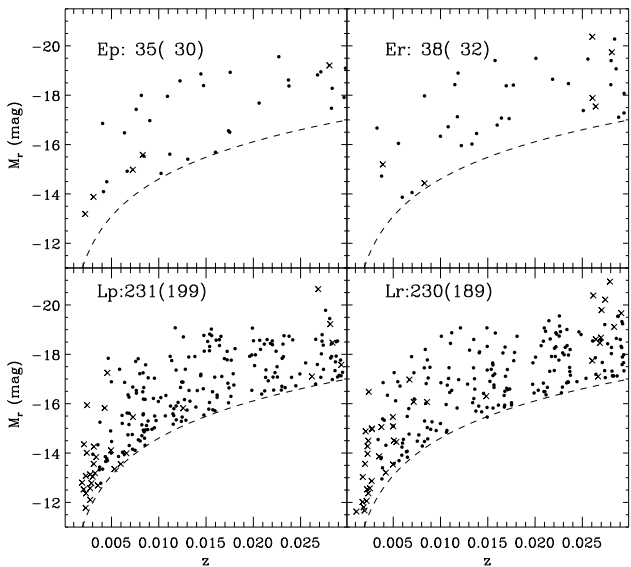}
\caption{Absolute magnitude vs. redshift 
  for early-type satellites in prograde orbit (Ep),
  for early-type satellites in retrograde orbit (Er),
  for late-type satellites in prograde orbit (Lp), and 
  for late-type satellites in retrograde orbit (Lr).
Satellites of the host galaxies in the luminosity subsample ($-18.5\geq M_{r,\rm host}>-21.5$)
  are filled circles, while the rests are crosses. 
Number outside the parenthesis means the number of galaxies in each panel,
  and the number in the parenthesis indicates the number of galaxies 
  within the luminosity subsample.
}\label{fig-sat}
\end{figure}

We plot $r$-band absolute magnitudes of host galaxies and their satellites
  as a function of redshift in Figures \ref{fig-host} and \ref{fig-sat}.
We made a luminosity subsample of satellites
  whose hosts have absolute magnitudes in the range of $-18.5\geq M_r>-21.5$.

\begin{figure}
\center
\includegraphics[width=85mm]{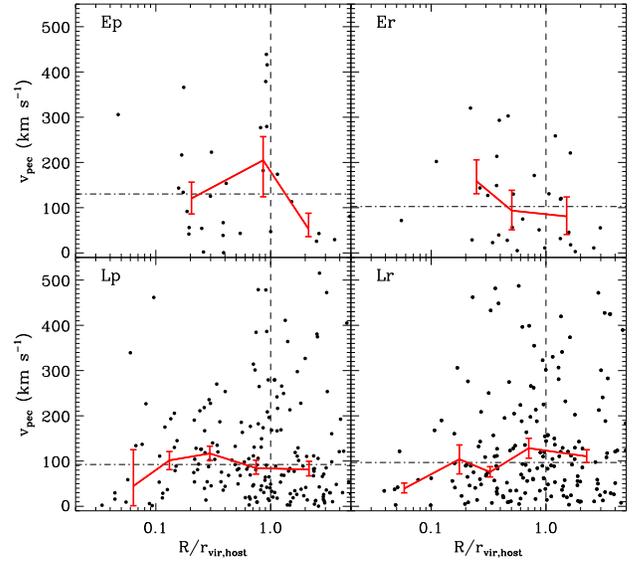}
\caption{Peculiar velocity $v_{\rm pec}$ vs. projected distance to the host galaxy
  for early-type satellites in prograde orbit (Ep),
  for early-type satellites in retrograde orbit (Er),
  for late-type satellites in prograde orbit (Lp), and 
  for late-type satellites in retrograde orbit (Lr)
  in the luminosity subsample.
Global median values are shown by horizontal dot-dashed lines, and
  the median value in each radial bin is connected by the solid line.
Vertical dashed line indicates one virial radius of the host.}\label{fig-vpec}
\end{figure}

In Figure \ref{fig-vpec}, we plot peculiar velocities of satellites 
  as a function of a distance to host galaxies.
It shows that global median values of $v_{\rm pec}$
  do not significantly vary depending on the orbit or morphology.
However, there exists an interesting trend
  that $v_{\rm pec}$ of the late-type satellites in prograde orbit (Lp) 
  increases as they approach the host, 
  while $v_{\rm pec}$ of those in retrograde orbit (Lr) decreases.
At the merger scales with $R/r_{\rm vir,host}<0.1$ the orbital velocity
  decreases for both Lp and Lr cases.

\section{Results}\label{results}
\subsection{Spatial Distribution of Satellites}

\begin{figure}
\center
\includegraphics[width=85mm]{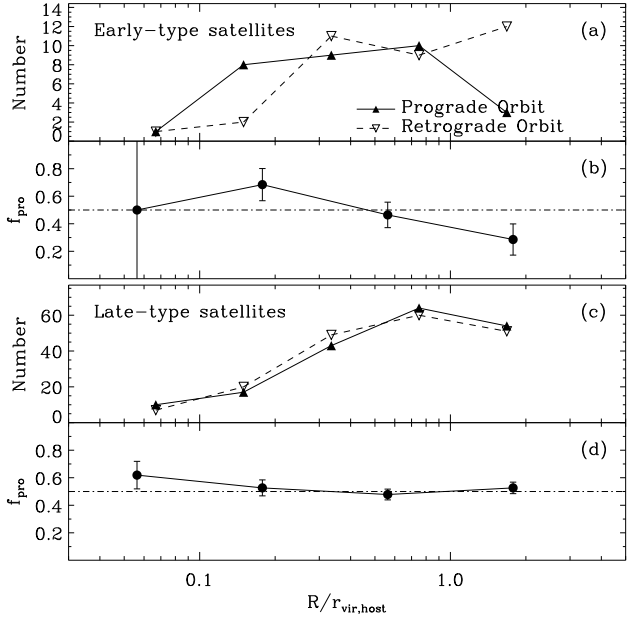}
\caption{Distribution of early-type satellites
  in prograde (filled triangles) and retrograde (open upside-down triangles) orbits ({\it a})
  and the number fraction of satellites in prograde orbit ($f_{\rm pro}$) 
  among early-type satellites ({\it b}) 
  as a function of the projected distance to the host galaxies.
Distribution of late-type satellites
  in prograde (filled triangles) and retrograde (open upside-down triangles) orbits ({\it c})
  and $f_{\rm pro}$ among late-type satellites ({\it d}).
}\label{fig-frac}
\end{figure}

\begin{figure}
\center
\includegraphics[width=85mm]{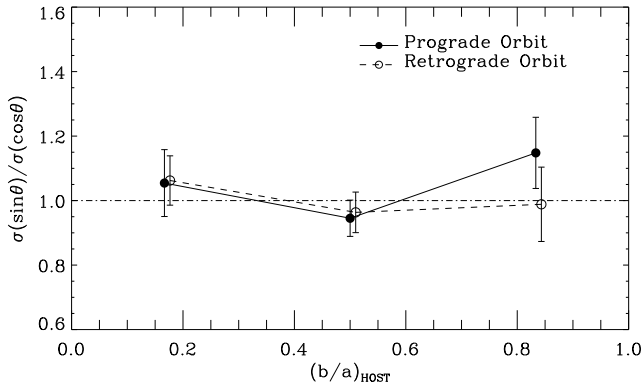}
\caption{Ratio of the dispersions of satellite distribution along 
  ($\sigma$(cos$\theta$)) and perpendicular ($\sigma$(sin$\theta$)) to
  the major axis of hosts as a function of the host axis ratio.
Solid and dashed lines indicate satellites in prograde and retrograde orbits, respectively.
}\label{fig-prob}
\end{figure}

In Figure \ref{fig-frac}, we plot the radial distribution of satellites
  in prograde and retrograde orbits.
We also show the fraction of galaxies in prograde orbit ($f_{\rm pro}$)
  as a function of the projected distance to host galaxies.
The top panel shows that
  satellites in prograde orbits are located closer to their hosts compared to those
  in retrograde orbits for the case of early-type satellites.
However, we do not find such dependence of the spatial distribution of late-type satellites
  on their orbits (see panel c).
As a result, $f_{\rm pro}$ of early types increases as $R$ decreases,
  but $f_{\rm pro}$ of late types remains constant ($\sim0.5$) within the error bar.

To study whether or not satellites are concentrated on the plane of host galaxies' disk, 
  we plot the ratio of the dispersion of satellite distribution along 
  ($\sigma$(cos$\theta$)) and perpendicular ($\sigma$(sin$\theta$)) to
  the major axis of the host
  as a function of the ratio of minor axis to the major axis of the host ($(b/a)_{\rm HOST}$)
  in Figure \ref{fig-prob}.
It shows neither of the distributions of prograde or retrograde orbit satellites 
  is aligned with late-type host galaxies.

\subsection{Luminosity}

\begin{figure}
\center
\includegraphics[width=85mm]{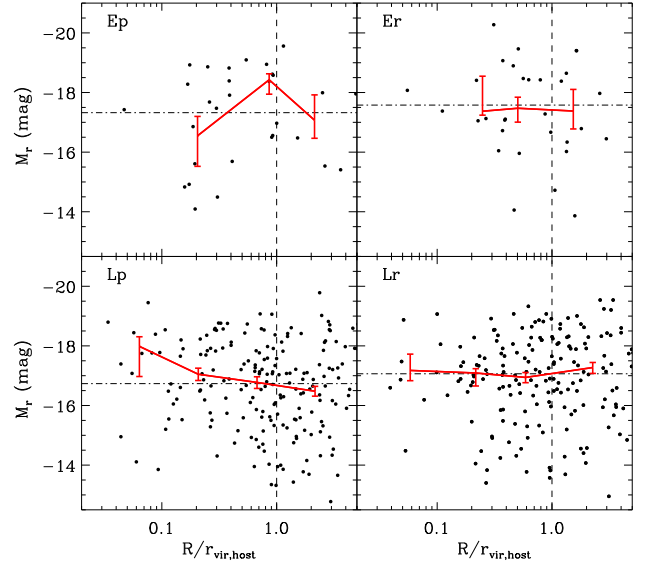}
\caption{Same as Fig. \ref{fig-vpec}, but for absolute magnitude $M_r$.
  }\label{fig-mag}
\end{figure}

Figure \ref{fig-mag} shows absolute magnitudes of the satellites 
  in the luminosity subsample
  as a function of the projected distance to host galaxies.
Each panel represents different subsamples depending on morphology and orbit.
For both early and late types,
  galaxies in retrograde orbit, on average, appear to be slightly brighter 
  than those in prograde orbit.
A statistical test with Monte Carlo simulation 
  to compute the significance of the difference in the median of the physical quantity   
  (to be discussed in the end of this section)
  confirms this difference for only late types with a significance level of 97$\%$.
When we compare satellites in each radial bin,
  the Monte Carlo test indicates with a significance level of 99$\%$ 
  that median absolute magnitudes of late-type galaxies 
  in the outer region ($R>r_{\rm vir,host}$) are different depending on the orbit.  
  
\begin{figure}
\center
\includegraphics[width=85mm]{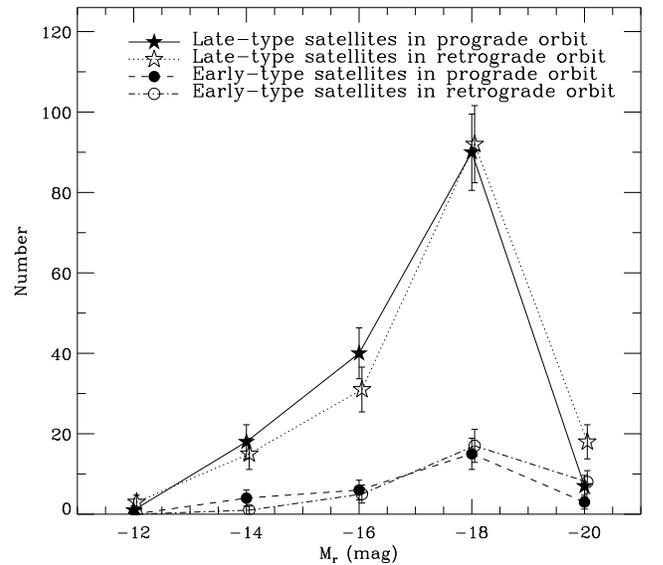}
\caption{Distribution of absolute magnitudes for all satellite galaxies.
Solid and dotted lines, respectively, indicate luminosity functions (LFs) 
  for late-type satellites in prograde and retrograde orbits.
Dashed and dot-dashed lines denote LFs for 
  early-type satellites in prograde and retrograde orbits, respectively.
}\label{fig-lum}
\end{figure}

To investigate the difference in the galaxy luminosity
  depending on the orbit in more detail,
  we show the distribution of absolute magnitudes in Figure \ref{fig-lum}.
We count the number of satellites within the redshift limit defined by 
  the apparent magnitude limit in Figure \ref{fig-sat}, and
  do not correct the count for different volumes.
It is seen that bright satellites ($-18\geq M_r>-20$) in retrograde orbit  
  are more numerous compared to those in prograde orbit.
The test with Monte Carlo simulation confirms this result with a significance level of 97$\%$
  in the sense that median absolute magnitudes of late types in prograde and retrograde orbits 
  are significantly different.

We summarize the results of statistical comparison of several physical parameters
  between satellites in prograde and retrograde orbits and
  between satellites in the inner and outer regions 
  in Tables \ref{tab-all} and \ref{tab-inout}, respectively. 

The statistical significance of the difference in the median or the dispersion of physical quantities,
  is computed by Monte Carlo simulations.
For a given physical parameter,
  we construct two subsamples (with the same number of galaxies as in the actual
  prograde and retrograde samples)
  by randomly drawing the observed parameters from the whole early-type or late-type 
  satellite galaxy sample
  and compute the median and the dispersion of each subsample.
The resulting two subsamples will have the same medians and standard deviations on average.
These random subsamples will tell us about whether or not
the difference in the parameter between prograde and retrograde satellite samples
  is statistically significant. 
We construct 1000 trial data sets and compute the fraction of simulated data sets 
  in which the difference of the physical quantity is larger than 
  or equal to that based on the real data (f$_{{\rm sim}\geq {\rm obs}}$).
The significance levels of the difference defined by 100(1$-$f$_{{\rm sim}\geq {\rm obs}}$) (\%) are given in Tables.
We have applied the statistical tests to all the physical parameters, 
  but only show statistically significant results (significance level above 95$\%$) in the Tables.

\subsection{Star Formation Activity Parameters}\label{sfa}

\begin{table*}
\begin{center}
\caption{Statistical Comparison between Satellites in Prograde and Retrograde Orbits\label{tab-all}}
\begin{tabular}{ccccccccc}
\hline\hline 
Parameter & Type & MC$_{\rm m,all}$ & MC$_{\rm d,all}$ &
 MC$_{\rm m,in}$ & MC$_{\rm d,in}$ &
 MC$_{\rm m,out}$ & MC$_{\rm d,out}$ & Remarks \\
  (1) & (2) & (3) & (4) & (5) & (6) & (7) & (8) & (9) \\
\hline
  $M_r$        & L & {\bf 97} & 58 & 49 & 40 &{\bf 99} & 69 &       Fig. \ref{fig-mag} \\
  W(H$\alpha$) & L & 56 & 49 & 33 & 11 & 89 & {\bf 97} &       Fig. \ref{fig-halpha} \\    
    $\Delta(g-i)$& L & {\bf 95} & {\bf 98} & 66 & 72 & {\bf 98} & {\bf 98} &       Fig. \ref{fig-gigrad} \\
  c$_{\rm in}$ & L & 38 & {\bf 98} & 47 & 88 & 38 & {\bf 97} &       Fig. \ref{fig-cin} \\
  \hline
\end{tabular}
\begin{flushleft}
{\it Column descriptions}.
(1): Physical parameters.
(2): Morphological types (E: early types, L: late types).
(3): Significance level ($\%$) of the difference of the median of the physical parameter
  between satellites in prograde and retrograde orbits
  determined from Monte Carlo simulation.
(4): Significance level ($\%$) of the difference of the dispersion of the physical parameter
  between satellites in prograde and retrograde orbits
  determined from Monte Carlo simulation.
(5): Same as column (3), but for satellites in the inner region ($R<1r_{\rm vir,host}$).
(6): Same as column (4), but for satellites in the inner region.
(7): Same as column (3), but for satellites in the inner region.
(8): Same as column (4), but for satellites in the outer region ($R\geq1r_{\rm vir,host}$).
(9): Relevant figure.
Statistically significant values ($\geq95\%$) are represented in bold face.
\end{flushleft}
\end{center}
\end{table*}

\begin{table}
\begin{center}
\caption{Statistical Comparison between Satellites in the inner and outer regions\label{tab-inout}}
\begin{tabular}{ccccccc}
\hline\hline 
Parameter & Type &  MC$_{\rm m,pro}$ & MC$_{\rm d,pro}$ &
 MC$_{\rm m,ret}$ & MC$_{\rm d,ret}$& Remarks \\
  (1) & (2) & (3) & (4) & (5) & (6) & (7)   \\
\hline
  ($u-r$)      & L & {\bf 99} & 67 & 88 & 65 &       Fig. \ref{fig-ur} \\
  W(H$\alpha$) & L & 13 & 12 & 64 & {\bf 97} &       Fig. \ref{fig-halpha} \\
  $\Delta(g-i)$& E & {\bf 96} & 61 &  4 & 43 &       Fig. \ref{fig-gigrad} \\
    $R_{\rm Pet}$& E & {\bf 97} & 77 & 90 & 24 &       Fig. \ref{fig-rpet} \\
  ($u-r$)      & L & {\bf 99} & {\bf 98} & 11 & 27 &       Fig. \ref{fig-urspec} \\
  ($u-r$)      & L & 93 & {\bf 98} & 90 & 67 &       Fig. \ref{fig-urearly} \\
\hline
\end{tabular}
\begin{flushleft}
{\it Column descriptions}.
(1): Physical parameters.
(2): Morphological types (E: early types, L: late types).
(3): Significance level ($\%$) of the difference of the median of the physical parameter
  of prograde orbit satellites in the inner and outer regions 
  determined from Monte Carlo simulation.
(4): Significance level ($\%$) of the difference of the dispersion of the physical parameter
  of prograde orbit satellites in the inner and outer regions 
  determined from Monte Carlo simulation.
(5): Same as column (3), but for retrograde orbit satellites.
(6): Same as column (4), but for retrograde orbit satellites.
(7): Relevant figure.
Statistically significant values ($\geq95\%$) are represented in bold face.
\end{flushleft}
\end{center}
\end{table}

\begin{figure}
\center
\includegraphics[width=85mm]{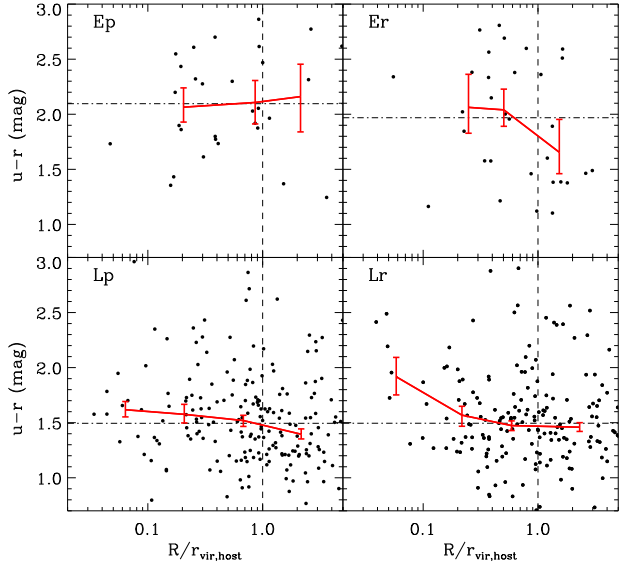}
\caption{Same as Fig. \ref{fig-vpec}, but for ($u-r$) color.}\label{fig-ur}
\end{figure}

Figure \ref{fig-ur} shows ($u-r$) colors of satellite galaxies
  divided by their morphologies and orbital motion as a function of 
  the projected distance to the hosts.
The late-type satellites in prograde and retrograde orbits have a similar global median value,
  and both show a hint of an increase in ($u-r$) color with decreasing $R$.
In the Monte Carlo test for the radial gradient we divide each sample into 
inner ($R< r_{\rm vir,host}$) and outer ($R\geq r_{\rm vir,host}$) regions,
and test if the difference in the physical parameter in two regions is statistically
significant. The random subsamples are generated in such a way that a parameter
value is randomly drawn from the combined sample and assigned to a galaxy in
inner or outer region. The resulting subsamples will have no radial gradient
in the parameter and have the same dispersions on avergae.
A test with such Monte Carlo simulation
  confirms that the change of ($u-r$) color with $R$ for late types in prograde orbit 
  (see Tab. \ref{tab-inout}) is significant at a 99\% confidence level.
It is noted that the increase of ($u-r$) colors for those in retrograde orbit
  is very strong in the innermost region ($R<0.1r_{\rm vir,host}$) even though
  the overall gradient is not signifiant.

\begin{figure}
\center
\includegraphics[width=85mm]{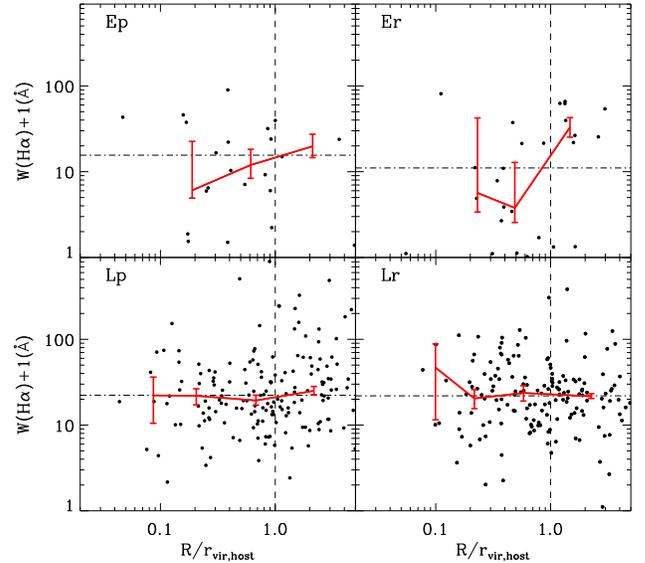}
\caption{Same as Fig. \ref{fig-vpec}, but for equivalent width of H$\alpha$ line.}
\label{fig-halpha}
\end{figure}

In Figure \ref{fig-halpha}, we plot the equivalent width of H$\alpha$ lines of satellite galaxies,
  which is a useful measure of star formation activity (SFA).
For late-type satellites,
  the median values of W(H$\alpha$) for two subsamples are similar,
  and those do not seem to change with $R$ for both samples
  even though their ($u-r$) colors become redder at smaller $R$ as seen in Figure \ref{fig-ur}.
However, the test with Monte Carlo simulation shows that
  the difference in the dispersion of W(H$\alpha$) 
  between late-type satellites in prograde and retrograde orbits
   is remarkable with a significance level of of 97\%  (see Tab. \ref{tab-all}).
The difference in the dispersion of W(H$\alpha$) 
  of retrograde orbit satellites in the inner and outer regions 
  is also noted (see Tab. \ref{tab-inout}).

\begin{figure}
\center
\includegraphics[width=85mm]{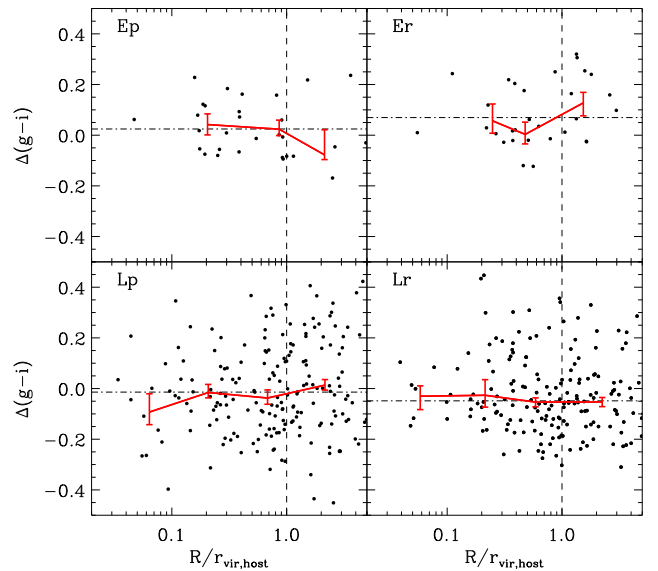}
\caption{Same as Fig. \ref{fig-vpec}, but for ($g-i$) color gradient.}
\label{fig-gigrad}
\end{figure}

Figure \ref{fig-gigrad} represents ($g-i$) color gradients of satellite galaxies.
The positive $(g-i)$ color gradient means
  a bluer color in the central region of a galaxy than that in the outer region.
For early types,
   the difference of median ($g-i$) color gradients of satellites in prograde orbit (Ep)
   between in the inner and outer regions is noticeable,
    which is supported by the test with Monte Carlo simulation (see Tab. \ref{tab-all}).
For late types,
  the test with Monte Carlo simulation  suggests that
  the median and its dispersion of ($g-i$) color gradients
  of satellites in prograde and retrograde orbits
  are different significantly in the whole/outer region.

\subsection{Structure Parameters}

\begin{figure}
\center
\includegraphics[width=85mm]{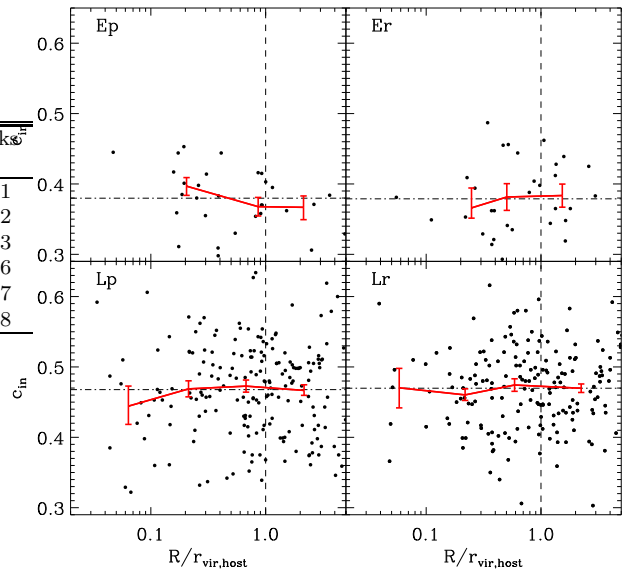}
\caption{Same as Fig. \ref{fig-vpec}, but for concentration index $c_{\rm in}$.}
\label{fig-cin}
\end{figure}

Figure \ref{fig-cin} shows the distribution of
  the concentration index $c_{\rm in}$ for satellite galaxies
  divided by their morphologies and orbital motion.
Small value of $c_{\rm in}$ means that the light of a galaxy 
  is more concentrated in the central region.
It is found that the global median and the scatter of $c_{\rm in}$ values
  for early-type satellites are not significantly different
  depending on their orbits.
However, it is noted that there are few early types in retrograde orbit
  with large $c_{\rm in}$ values ($>0.4$) in the inner region ($R<0.2r_{\rm vir,host}$),
  and few early types in prograde orbit with large $c_{\rm in}$ values ($>0.4$) 
  are found in the outer region ($R>r_{\rm vir,host}$).
For late types,
  it is seen that the difference in the dispersion of $c_{\rm in}$
  between late-type satellites in the prograde and retrograde orbits (see Tab. \ref{tab-all}).
However, the global median values of $c_{\rm in}$ for two subsamples are similar,
  and the median values for both subsamples do not show any significant variation with $R$.

\begin{figure}
\center
\includegraphics[width=85mm]{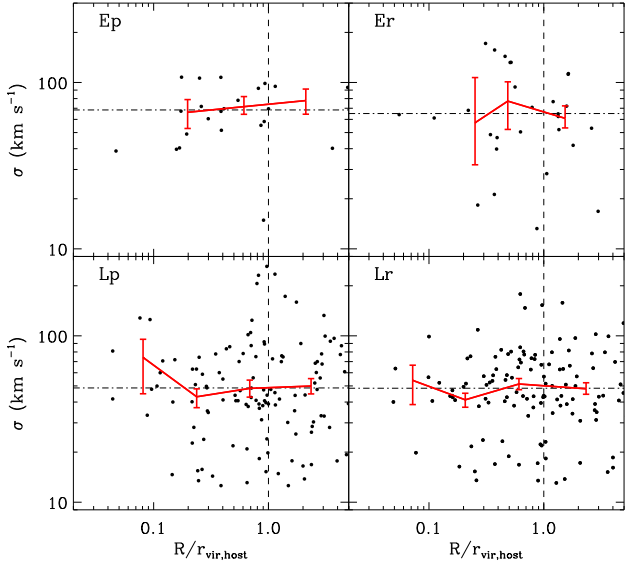}
\caption{Same as Fig. \ref{fig-vpec}, but for central velocity dispersion $\sigma$.
}\label{fig-vdisp}
\end{figure}

\begin{figure}
\center
\includegraphics[width=85mm]{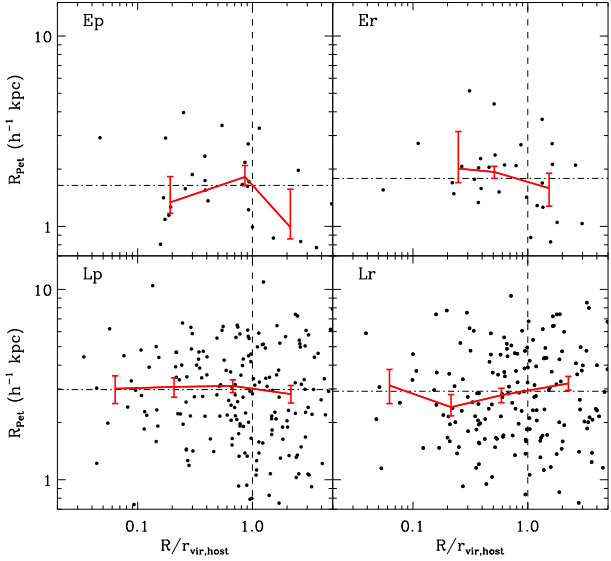}
\caption{Same as Fig. \ref{fig-vpec}, but for Petrosian radius $R_{\rm Pet}$.
}\label{fig-rpet}
\end{figure}

In Figure \ref{fig-vdisp}, we plot
 central velocity dispersions of satellite galaxies.
The median values for early types in prograde and retrograde orbits
  are similar and do not show any significant variation with $R$.
For late types,
  any significant difference 
  between two subsamples of late-type satellites
  is not found except for the scatter of $\sigma$ values.

Figure \ref{fig-rpet} represents the $i$-band Petrosian radius 
  $R_{\rm Pet}$ of satellite galaxies.
The $R_{\rm Pet}$ for the early-type satellites
  in the inner region ($R<r_{\rm vir,host}$) 
  appears to be larger than that in the outer region ($R>r_{\rm vir,host}$) (see Tab. \ref{tab-inout}),
  while no significant difference between the late-type satellites 
  in prograde and retrograde orbits is not seen.

\section{Discussion}\label{discuss}

\subsection{Different Galaxy Samples}\label{samp}

\begin{figure}
\center
\includegraphics[width=85mm]{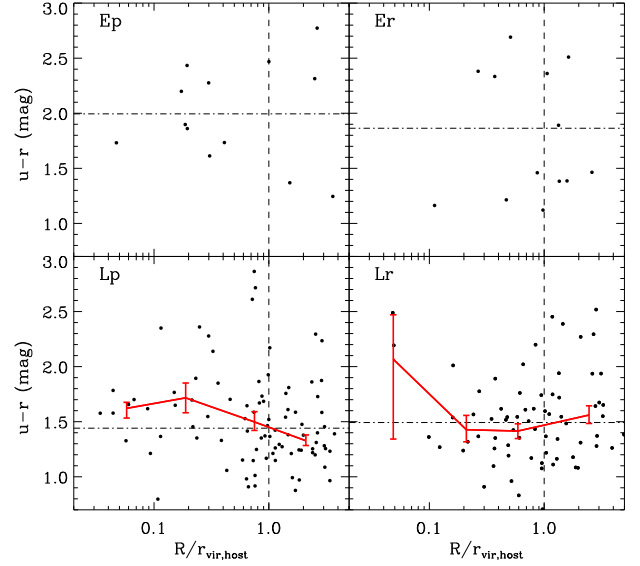}
\caption{Same as Fig. \ref{fig-ur}, 
  but for the satellites associated to host galaxies whose 
  spin direction is determined from the spectroscopic observations.}\label{fig-urspec}
\end{figure}

\begin{figure}
\center
\includegraphics[width=85mm]{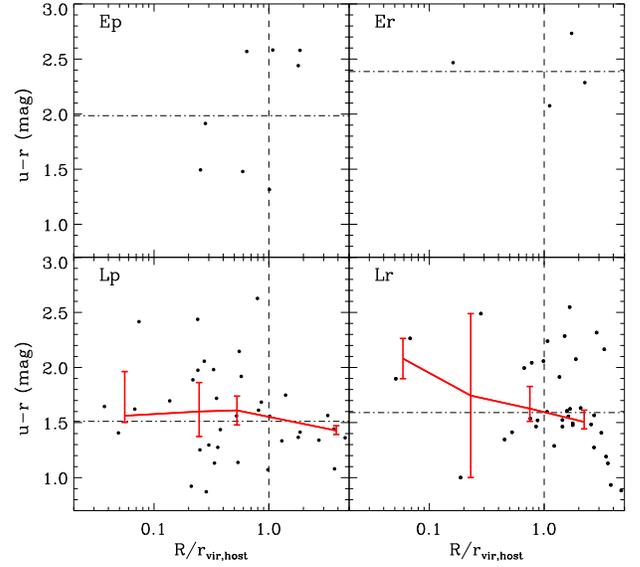}
\caption{Same as Fig. \ref{fig-ur}, 
  but for the satellites associated to early-type hosts.}\label{fig-urearly}
\end{figure}

We found in Figures \ref{fig-halpha}-\ref{fig-gigrad}
  and Table \ref{tab-all} that
  the median or its dispersion of W(H$\alpha$) and ($g-i$) color gradients
  can be different depending on the orbital motion 
  of late-type satellites.
The SFA of early-type satellites in retrograde orbit (Er)
  also appears to be stronger than those in prograde orbit 
  in the outer region ($r_{\rm vir,host}<R$) as seen 
  in Figures \ref{fig-ur} and \ref{fig-halpha}.
Moreover, ($u-r$) colors of late-type satellites
  increase as they approach their host galaxies
  for both satellites in prograde and retrograde orbits,
  and ($u-r$) colors of late-type satellites in retrograde orbit 
  in the innermost region ($R<0.1r_{\rm vir,host}$) are remarkably large.
These differences depending on the orbit can be attributed to 
  the different origin of satellite galaxies,
  or to the different strength of physical processes that they experience through
  interactions with host galaxies.

To check whether or not these results are strongly affected by the satellites
  associated to the host galaxies whose spin direction is determined 
  from the SDSS color images in this study,
  in Figure \ref{fig-urspec}, 
  we plot ($u-r$) colors versus the projected distance to the hosts
  using the satellites associated to the host galaxies whose spin direction is determined 
  from the spectroscopic observations.
For both early and late types,
  we do not find any significant difference of ($u-r$) colors
  between satellites in prograde and retrograde orbits,
  which is consistent with the result 
  when we use the combined sample of satellites
  in the host galaxies whose spin direction is determined 
  from the color images or the spectroscopic observations (see Fig. \ref{fig-ur}).
In addition, the late-type satellites in prograde orbits (Lp) show an increase in 
  ($u-r$) color as decreasing $R$, which is again similar to the case of
  late types in prograde orbits using the combined sample of satellites
  (see Lp in Fig. \ref{fig-ur} and Tab. \ref{tab-inout}).
In the result, the results of SFA using the combined sample of satellites
  are not affected strongly by the satellites
  associated to the host galaxies whose spin direction is determined 
  from the color images in this study.

Since the physical parameters of satellite galaxies can be strongly affected 
  by the morphology of host galaxies (e.g., \citealt{ann08}),
  it is also interesting to compare satellite systems
  that have early- and late-type host galaxies.
To investigate the SFA of satellites around early-type hosts
  we use a sample of 101 satellites associated with 24 early-type hosts,
  and show their ($u-r$) colors in Figure \ref{fig-urearly}.
It is seen that ($u-r$) colors of late-type satellites become redder as
  $R$ decreases, which was also seen in Figure \ref{fig-ur} or \ref{fig-urspec}.
For early-type satellites,
  the small number of satellites prevents us from drawing any firm conclusion.

\subsection{Comparison with Other Galaxy Systems}

Previously, \citet{park08} and \citet{pc09} 
  found that galaxy properties strongly depend on the
  distance and the morphology of the nearest neighbor galaxy.
For example, the SFA of galaxies is enhanced as they approach late-type neighbors
  of comparable luminosity,
  while it is reduced as they approach early-type neighbors.
The structure parameters of galaxies change with 
  the distance to the nearest neighbor galaxy
  in the same direction independently of neighbor's morphology.
Galaxy properties start to change when galaxies approach
  their nearest neighbor galaxy closer than
  the virial radius ($r_{\rm vir,nei}$) of the nearest neighbor galaxy and
  to change significantly at $R< 0.05r_{\rm vir,nei}$ 
  where the galaxies in pair start to merge.
They interpreted this phenomenon as a result of hydrodynamic interactions
  within the virial radius of the galaxy plus dark halo system.

The dependence of satellite properties as a function of
  the distance to the host galaxy in this study can be compared
  with the case of late-type neighbors in \citet{pc09}.
Since we restrict our analysis to the case of late-type host galaxy,
  it is expected that the SFA of both early- and late-type satellites
  increases as they approach the host.
However, we found in \S \ref{sfa} that ($u-r$) color becomes redder (Fig. \ref{fig-ur}) 
  and $W(H\alpha$) hardly changes (Fig. \ref{fig-halpha}) 
  even when late-type satellites approach their late-type hosts.
This may be because, unlike the galaxies considered by \citet{pc09}, 
  the satellites in this study are much fainter and less massive than their hosts. 
As a result they may not be able to pick up cold gas from their host much,
  and the enhancement of SFA through interactions with their host is counterbalanced
  by the quenching by the hot halo gas of the host.

On the other hand, the environment of satellite galaxies
  can be compared with that of galaxies in galaxy clusters
  because the member galaxies in a galaxy cluster are bounded within the gravitational 
  potential well of the cluster as the satellites are bounded by their host galaxy.
\citet{ph09} studied the dependence of galaxy properties on the clustercentric radius 
  and the environment attributed to the nearest neighbor galaxy 
  using the SDSS galaxies associated with the Abell galaxy clusters.
They found that the SFA of late-type galaxies decreases 
  as they approach the cluster center,
  and suggested that hydrodynamic interactions with
  nearby early-type galaxies in clusters
  is the main drive to quenching the SFA of late types.
Most cluster galaxies are early types including the brightest cluster galaxies 
 (or cD galaxies), which might be comparable with the case of
  early-type host galaxy in the satellite system as seen in Figure \ref{fig-urearly}.
The late-type satellites (bottom panels in Fig. \ref{fig-urearly})
  become redder as they approach early-type host galaxies,
  which agrees to the case of galaxy clusters.

Previously, \citet{zinn93} pointed out that there are two distinct populations
  in the halo globular clusters in the Milky Way: old and young halo clusters.
Two populations were originally divided by the horizontal branch morphology,
  but were found to be also separated by their orbital motion in the Milky Way:
  prograde orbit for old halo clusters and 
  retrograde orbit for young halo clusters \citep{zinn93,van93,lee07}.
Therefore, the formation history of old halo clusters with prograde orbit
  is thought to be different from that of young halo clusters
  with retrograde orbit (e.g., \citealt{els62} vs. \citealt{sz78}).
For example, it was suggested that the clusters in retrograde orbit 
  may have been accreted into the Milky Way (e.g., \citealt{din99}),
  and further that these clusters and Galactic dSphs may 
  have the same origin in accretion or in partially disruption of
  formerly large parent satellite galaxies (e.g., \citealt{vm04}). 
Therefore, two populations of satellite galaxies with different orbits 
  in this study are also expected to have formed and evolved in different paths,
  which may be responsible for the different SFA depending on the orbit. 
The Galactic satellite galaxies also can give us important clues 
  for testing this scenario, but it is not achievable at this moment 
  because of small number ($\sim10$) of Galactic satellites 
  whose orbital motion is determined.

\section{Conclusions}\label{sum}

We have studied the orbital dependence of various galaxy properties 
  in the satellite systems of galaxies in SDSS.
Our primary results are summarized as follows.

\begin{enumerate}
\item We have found the satellite systems of galaxies at $z<0.03$,
  which consist of 8904 satellites associated with 3515 isolated host galaxies.

\item We have determined the spin direction of late-type host galaxies
  by inspecting the SDSS color images.
By combining our results with those 
  from the previous spectroscopic observations in the literature,
  we obtained 635 satellites associated with 215 host galaxies
  whose spin directions are known. 

\item The number ratio of satellites in prograde orbit to those in retrograde orbit
  is nearly equal to one, and the peculiar velocity distribution is found to be symmetric.
The number ratio of satellites in prograde orbit to those in retrograde orbit ($f_{\rm pro}$)
  appears to increase as the projected distance to the host galaxy decreases.
However, we do not find any significant difference in the azimuthal distribution 
  of satellites between those in prograde and retrograde orbits.

\item The satellites in retrograde orbit, on average, appear to be slightly 
  brighter than those in prograde orbit.
The latter seems to get brighter as they approach late-type hosts,
  while the former does not show any significant change with the distance to the hosts.

\item For early-type satellites in retrograde orbit,
  the SFA seems to decrease as they approach host galaxies.
For late-type satellites, ($u-r$) colors of both samples 
  in prograde and retrograde orbits
  appear to increase as they approach the hosts,
  while the radial variations of 
  the equivalent width of H$\alpha$ lines for both samples are not significant.

\item The structure parameters of satellites
  do not seem to be different significantly between galaxies in prograde and retrograde orbits.

\end{enumerate}

\acknowledgments

The authors are grateful to the anonymous referee for useful
  comments that improved the original manuscript.
CBP acknowledges the support of 
the National Research Foundation of Korea(NRF) grant funded by the 
Korea government(MEST) (No. 2009-0062868).
Funding for the SDSS and SDSS-II has been provided by the Alfred P. Sloan 
Foundation, the Participating Institutions, the National Science 
Foundation, the U.S. Department of Energy, the National Aeronautics and 
Space Administration, the Japanese Monbukagakusho, the Max Planck 
Society, and the Higher Education Funding Council for England. 
The SDSS Web Site is http://www.sdss.org/.
The SDSS is managed by the Astrophysical Research Consortium for the 
Participating Institutions. The Participating Institutions are the 
American Museum of Natural History, Astrophysical Institute Potsdam, 
University of Basel, Cambridge University, Case Western Reserve University, 
University of Chicago, Drexel University, Fermilab, the Institute for 
Advanced Study, the Japan Participation Group, Johns Hopkins University, 
the Joint Institute for Nuclear Astrophysics, the Kavli Institute for 
Particle Astrophysics and Cosmology, the Korean Scientist Group, the 
Chinese Academy of Sciences (LAMOST), Los Alamos National Laboratory, 
the Max-Planck-Institute for Astronomy (MPIA), the Max-Planck-Institute 
for Astrophysics (MPA), New Mexico State University, Ohio State University, 
University of Pittsburgh, University of Portsmouth, Princeton University,
the United States Naval Observatory, and the University of Washington. 

{}
\end{document}